\documentclass[12pt,english]{article}
\usepackage{babel}
\usepackage{epsfig}
\usepackage{amsmath}
\usepackage{amssymb} 
\usepackage[small]{caption2} 
\usepackage{fleqn} 
\usepackage{graphicx} 
\usepackage[small,loose]{subfigure}  
\usepackage{cite} 
\advance \headheight by 3.0truept       
\setcaptionwidth{.85\textwidth}

\newcommand{\CenterObject}[1]{\ensuremath{\vcenter{\hbox{#1}}}}

\makeatother

\newcommand{\lwig}{\mbox{\;\raisebox{.3ex}
    {$<$}$\!\!\!\!\!$\raisebox{-.9ex}{$\sim$}\;}}
\newcommand{\gwig}{\mbox{\;\raisebox{.3ex}
    {$>$}$\!\!\!\!\!$\raisebox{-.9ex}{$\sim$}}\;}

\begin{document}
\title{
\begin{flushright}
{\normalsize  IPPP/05/54} 
\end{flushright}
\vspace{1cm}
{\bf Neutron--Electron EDM Correlations in Supersymmetry
and Prospects for EDM Searches}\\[1cm]}
\author{%
\textbf{Steven A. Abel }\\[0.4cm]
{\normalsize\textit{$$IPPP, University of Durham, DH1 3LE Durham, UK}}\\[0.4cm]
\textbf{Oleg Lebedev }\\[0.4cm]
{\normalsize\textit{$$Deutsches Elektronen-Synchrotron DESY, 
22603 Hamburg, Germany}}}
\maketitle 
\hrule
\abstract{ Motivated by recent progress in experimental techniques of
electric dipole moment (EDM) measurements,
we study correlations between  the neutron and electron EDMs  in common supersymmetric models.
These include minimal supergravity (mSUGRA)  with small CP phases, 
mSUGRA with a heavy  SUSY spectrum,
the decoupling scenario and  split SUSY. 
In most cases, 
the electron and neutron EDMs are found to be observable in the next round of EDM experiments.
They exhibit certain correlation patterns. For example, if $d_n \sim 10^{-27 } \; e\;{\rm cm}$
is found,  $d_e$ is predicted to lie in the range  $ 10^{-28 }-10^{-29 }  \; e\;{\rm cm}$.
   }
\vspace{0.5cm}
\hrule
\vspace{1.5cm}

\thispagestyle{empty}

\newpage

\section{Introduction}

The current EDM limits \cite{Harris:1999jx} --\cite{Romalis:2000mg}\footnote{
The EDM collaboration has recently announced 
an improved (preliminary)  neutron EDM bound
$\vert d_n\vert \lwig 3\times 10^{-26} \; e\;{\rm cm}$ \cite{susy}.  }
\begin{eqnarray}
&& \vert d_n\vert < 6\times 10^{-26} \; e\;{\rm cm}\;, \nonumber\\
&& \vert d_e \vert < 2\times 10^{-27} \; e\;{\rm cm}\;, \nonumber\\
&& \vert d_{\rm Hg}\vert  < 2\times 10^{-28} \; e\;{\rm cm}
\label{bound}
\end{eqnarray}
impose severe constraints on CP violating physics beyond the Standard Model
and, in particular, supersymmetry. Supersymmetric models accommodating the above
bounds usually predict EDMs not far from the current limits which makes the next
round of EDM experiments with the sensitivity $d_n \sim 10^{-28 } \; e\;{\rm cm}$,
$d_e \sim 10^{-30 } \; e\;{\rm cm}$   \cite{susy} --\cite{lanl}   particularly interesting.
Other EDM experiments such as measurements of the deuteron and muon EDMs 
\cite{Semertzidis:2003iq}
would provide important complementary information on CP violation in supersymmetry.

In this work, we address the questions ``What are the  EDM expectations
in SUSY models ?'' and ``How can one distinguish different  sources of EDMs ?''.
It is well known that generic SUSY models predict too large EDMs
which constitutes the SUSY CP problem \cite{Ellis:1982tk}.
The problem is resolved  in certain classes of supersymmetric models, of which
we choose four representative types. These include SUSY models with

$\bullet$ small CP phases

$\bullet$ heavy spectrum

$\bullet$ decoupling

$\bullet$ split SUSY

To determine the source of EDMs, 
we study correlations between the neutron and electron electric dipole moments.
In particular, if the neutron EDM is due to the QCD $\theta$--term 
\cite{Baluni:1978rf} --\cite{Pospelov:1999ha}, one expects
very small leptonic EDMs. On the other hand, in supersymmetry
both the hadronic and leptonic EDMs are enhanced and 
there exist certain correlation patterns. Determination
of such patterns is the main subject of this work.

The relevant low energy Lagrangian describing interactions of electrons, quarks, gluons and photons
is given by
\begin{eqnarray}
\label{leff}
&{\cal{L}}& = \theta {\alpha_s \over 8 \pi} G \tilde G+
\frac{1}{3} w\  f^{a b c} G^{a} \widetilde{G}^{ b}
G^{ c}  -
\frac{i}{2} d_f\ \overline{ f} (F\sigma)\gamma_5 f  -
\frac{i}{2} 
 g_s\widetilde{d_q}\ \overline{ q}  (G\sigma)\gamma_5 q ~, \nonumber  
\end{eqnarray}
where $w$, $d_f$, $\tilde d_q$ are the Weinberg operator \cite{Weinberg:dx}
coefficient,
the fermion  EDM, and the quark chromo-EDM (CEDM), respectively. 
Here $F$ and $G$ represent the photon and gluon field strengths, respectively. 
The electron EDM is given by $d_e$, whereas 
the neutron EDM is a model--dependent function of $\theta$, $w$, $d_q$ and  $\tilde d_q$.
We will use the naive dimensional analysis (NDA)  approach \cite{Arnowitt:1990eh},\footnote{The QCD correction factor for the $d_{u,d}$ contributions has recently been
recalculated \cite{Degrassi:2005zd}   
and found to be a factor of 2 or so smaller than that in 
\cite{Arnowitt:1990eh}. This leads to somewhat smaller estimates for $d_n$. } 
\begin{equation}
 d_n^{\rm NDA} \sim 2 d_d - 0.5 d_u + e(0.4 \tilde d_d -0.1 
\tilde d_u ) + 0.3 ~{\rm GeV}\times e w ~,
\end{equation}
where we have set $\theta=0$. 
Here the Wilson coefficients $d_q, \tilde d_q, w$ are evaluated at the
electroweak scale. The chromo--EDM and the Weinberg operator contributions
involve considerable uncertainties.
Furthermore, the result is sensitive to the quark masses which
we choose as $m_u(M_Z)=2$ MeV and  $m_d(M_Z)=4$ MeV.
The approach based on QCD sum rules gives a somewhat similar 
result \cite{Pospelov:2000bw},
\begin{equation}
 d_n^{\rm SR} \simeq 2 d_d - 0.5 d_u +e( \tilde d_d +0.5 
\tilde d_u) + 0.1 ~{\rm GeV}\times e w ~\;. 
\end{equation}
In the  Weinberg operator contribution,
we have used the (model--dependent) renormalization factor of 
Ref.\cite{Arnowitt:1990eh}.
In practice,  the two approaches usually agree  within a factor of 2 
(unless $w$ dominates) which 
suffices for our purposes. 
We note that there are also  neutron models which include the strange quark contribution \cite{Ellis:1996dg}.
This effect is difficult to estimate and involves large uncertainties. We defer a
study of such models until a subsequent publication.

\section{EDMs in the Standard Model}

We start by considering EDMs induced by the QCD $\theta$--term.
This is the primary source of hadronic EDMs in the Standard Model \cite{Baluni:1978rf} --\cite{Pospelov:1999ha},\cite{Shintani:2005xg}.  
In particular \cite{Pospelov:1999ha},
\begin{equation}
d_n \simeq 3 \times 10^{-16}~\theta \; e\;{\rm cm}
\end{equation}
with about 50\% uncertainty. 
On the other hand, the electroweak contributions are very small, $d_n \sim  10^{-32}\; e\;{\rm cm} $ 
\cite{Gavela:1981sk}. To satisfy the experimental bound (\ref{bound}), the $\theta$ parameter has to be tiny,
$\leq {\cal O}(10^{-10})$. Such a small value can hardly be explained by the Standard Model,
which constitutes the ``strong CP problem''.
The most popular solution to this problem in extensions of the
SM  invokes an anomalous 
Peccei--Quinn symmetry \cite{Peccei:1977hh}, which sets $\theta$
to zero. However, this symmetry is expected to be broken by higher
dimensional operators generated at the Planck scale \cite{Holman:1992us},
such that the resulting $\theta$ is finite but small. 
There are also alternative solutions to the strong CP problem
which employ other symmetries \cite{Mohapatra:fy}. In these models,
a finite $\theta$  can be induced by
radiative corrections. In either case, a small neutron EDM is expected but it is
difficult to make any quantitative prediction.

The $\theta$--term also induces nuclear EDMs. 
In particular, it
generates EDMs of the deuteron and the mercury atom.
The former receives contributions from both the constituent nucleons
and nucleon interactions, and can be evaluated via QCD sum rules 
with about 50\% uncertainty
\cite{Lebedev:2004va}. An EDM of the mercury atom is induced by the
Schiff moment which appears due to  CP violating isoscalar and isovector
pion--nucleon couplings. Despite recent progress in evaluating these
contributions \cite{Dmitriev:2003kb}, 
there is still an order of magnitude uncertainty in $d_{\rm Hg}(\theta)$,
whereas $d_{\rm Hg}$ induced by  the quark (colour--) EDM  contributions
is understood much better. 
The result is
\begin{eqnarray}
 && d_D \simeq - 1 \times 10^{-16} ~\theta \; e\;{\rm cm} \;, \nonumber \\
 && \vert d_{\rm Hg} \vert \sim {\cal O}(10^{-18}-10^{-19})~ \theta \; e\;{\rm cm}. 
\end{eqnarray}
The above formulae together with suppressed leptonic EDMs provide a correlation pattern
for the $\theta$--induced electric dipole moments.

It is important to remember that $d_e$ is not measured directly, but instead  is derived
from atomic EDMs. In particular, the current EEDM bound is due to the thallium atom
EDM measurement \cite{Regan:ta} and the relation   $d_{\rm Tl}\simeq -585 ~d_e$. 
If the $\theta$--term is non--zero, this relation is altered. 
Although no reliable calculation of $d_{\rm Tl} (\theta)$
is available, rough estimates \cite{Khriplovich:ga} of the valence proton EDM
contribution    give $d_{\rm Tl} \sim 10^{-20}~\theta~e$ cm. 
On the other hand, the SM electroweak interactions (including Majorana neutrinos)
 usually induce $d_e$ of order $10^{-38}e$ cm  \cite{Archambault:2004td}. 
Thus, in the Standard Model with $\theta > 10^{-15}$, 
 $d_{\rm Tl}$ is dominated by the $\theta$--background and is  less 
sensitive to $d_e$. This illustrates  that the effects of the $\theta$--term
can be  important for atomic systems and should be taken into account.

\section{EDMs in supersymmetry}

In supersymmetric models, there are additional sources of CP violation
associated with complex phases in the SUSY breaking F--terms
and, in addition, flavour misalignment between the Yukawa matrices
and the soft breaking terms \cite{Abel:2001cv}. 
EDMs are generated already at the one loop level
and typically exceed the experimental bounds by orders of magnitude
\cite{Ellis:1982tk}. This constitutes the SUSY CP problem.
The problem is alleviated  in certain classes of supersymmetric models.
These include models with small SUSY CP phases, those with a heavy SUSY spectrum,
the decoupling scenario and split SUSY. In what follows, we study neutron--electron
EDM correlations in these types of models. 

We note that there are also other possibilities for EDM suppression.
For instance, CP violation may have flavour off--diagonal nature due to some symmetry
\cite{Babu:2001se},\cite{Abel:2000hn}. We defer a study of this option until a
subsequent publication.
Finally,   EDMs may be  suppressed due to accidental cancellations
among independent terms \cite{Falk:1996ni}.
This option is however disfavoured by the mercury 
EDM constraint \cite{Falk:1999tm} --\cite{abelprl}.

In our numerical analysis, we choose two representative values of
$\tan\beta$, 5 and 35, and analyze the NEDM--EEDM correlations separately for 
SUSY CP violation induced by the phase of the $\mu$--term, $\phi_\mu$, and CP violation
due to the phase of the trilinear A--terms, $\phi_A$.
We scan over the parameter space of a given model and present our results as scatter plots
$d_e$ vs $d_n$.

\begin{figure}[t]
 \centerline{\CenterObject{\includegraphics[width=6.0cm]{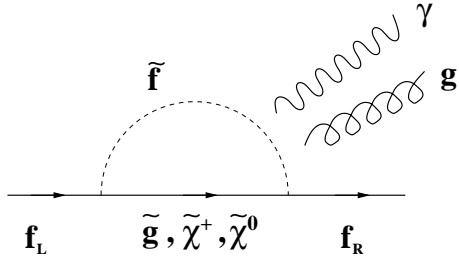}}}
\vspace*{0.3cm}
 \caption{
\footnotesize 
One loop SUSY contributions to fermion (C)EDMs. 
 }
\label{f1} 
\end{figure}

\subsection{Small CP phases}

First, we study the minimal SUGRA model (mSUGRA) with small 
($10^{-2}$) CP phases.
The smallness of the physical phases may  be due to approximate alignment
between the phases of the soft terms as  occurs in the dilaton--domination
scenario \cite{Kaplunovsky:1993rd}   
with the Giudice--Masiero mechanism for the $\mu$--term, 
or due to some approximate symmetry \cite{Lebedev:2002zt}.

The model is defined in terms of the following GUT scale parameters,
\begin{eqnarray}
&~~~~~~~~~~~~~~~~~~~~m_0~, ~m_{1/2}~,~\vert A \vert~, ~\tan\beta~& 
\label{msugra}
\end{eqnarray}
and CP--phases $\phi_A$, $\phi_\mu$. Here $m_0$ is the universal scalar mass,
$m_{1/2}$ is the universal gaugino mass and $A$ is the trilinear parameter.
The Higgs potential parameters $\vert \mu \vert$ and $\vert B\mu \vert$ are
found by imposing radiative electroweak symmetry breaking. We assume that  the
CP phases associated with the gaugino mass and the $B\mu$--term have been rotated
away by appropriate $U(1)_{\rm R}$ and $U(1)_{\rm PQ}$ transformations.
In this case, the physical CP phases are parametrized by $\phi_A$ and  $\phi_\mu$.

The (C)EDMs are 
dominated by one--loop diagrams involving gluinos, charginos and
neutralinos, Fig.\ref{f1},
\begin{eqnarray} 
&& d_q= d_q^{\tilde g} + d_q^{\tilde \chi^+} + d_q^{\tilde \chi^0}  \;,
\nonumber\\
&& \tilde d_q= \tilde d_q^{\tilde g} + \tilde d_q^{\tilde \chi^+} + 
\tilde d_q^{\tilde \chi^0}  \;,
\nonumber\\
&& d_e = d_e^{\tilde \chi^+} + d_e^{\tilde \chi^0} \;. 
\end{eqnarray}
The two loop contributions are considerably smaller.
The relevant formulae can be found, for example, in
 Ibrahim and Nath, Ref.\cite{Falk:1996ni}.

To get a feeling for  the size of the EDMs, let us consider a simple approximation 
$\phi_\mu \sim \phi_A \equiv \phi \ll 1$, $\tan\beta \sim 3$
and assume a single mass scale $M$ for the SUSY parameters (at the 
electroweak scale). Then,
\begin{eqnarray}
&& d_n \sim \left( {300 {\rm ~GeV} \over M} \right)^2 \sin\phi  \times 10^{-24} e\;{\rm cm}\;,\nonumber\\
&& d_e \sim \left( {300 {\rm ~GeV} \over M} \right)^2 \sin\phi  \times 10^{-25} e\;{\rm cm} \;.
\label{edm}
\end{eqnarray}
Both $d_n$ and $d_e$ grow linearly with $\tan\beta$.
Clearly, for the SUSY spectrum with electroweak masses, the CP phase has to be of order $10^{-2}$.
We see that the neutron and electron EDMs differ by about an order of magnitude, however
no prediction of their magnitudes can be made since the result depends on how small the phase is.
If no $d_e$ at the level  $10^{-30}~ e\;{\rm cm}$ is found, the CP phase will have to be smaller 
than $10^{-5}$ which appears highly unnatural. It would be rather difficult to engineer a robust
mechanism which would force the phases in the Lagrangian to align with such an extraordinary
precision. Thus, one may argue that $d_e$ and $d_n$ in this scenario
should not be far below the current experimental limits. 

In the same simple approximation, the mercury atom EDM is given by
$d_{\rm Hg} \sim (300~{\rm GeV}/M)^2 \sin\phi \times 10^{-26}~e$ cm.
Here $d_{\rm Hg}$ is dominated by the quark CEDMs \cite{Falk:1999tm}
and we have included an extra factor $\sim$1/4 due to a recent reevaluation
of the nuclear/atomic matrix elements. The deuteron EDM is expected to be similar
to the neutron EDM, $d_D \sim d_n$. 

\begin{figure}[t]
\centerline{\CenterObject{\includegraphics[width=5.5cm]{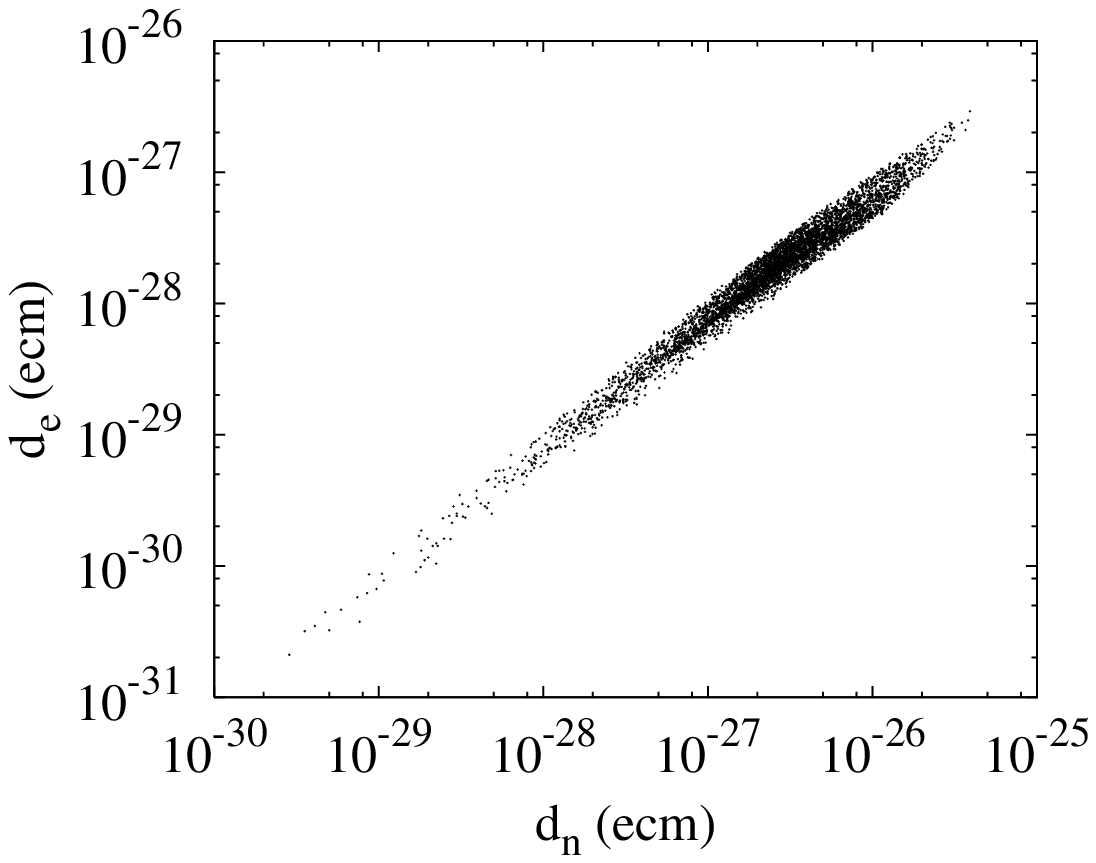} \hspace{0.5cm}
\includegraphics[width=5.5cm]{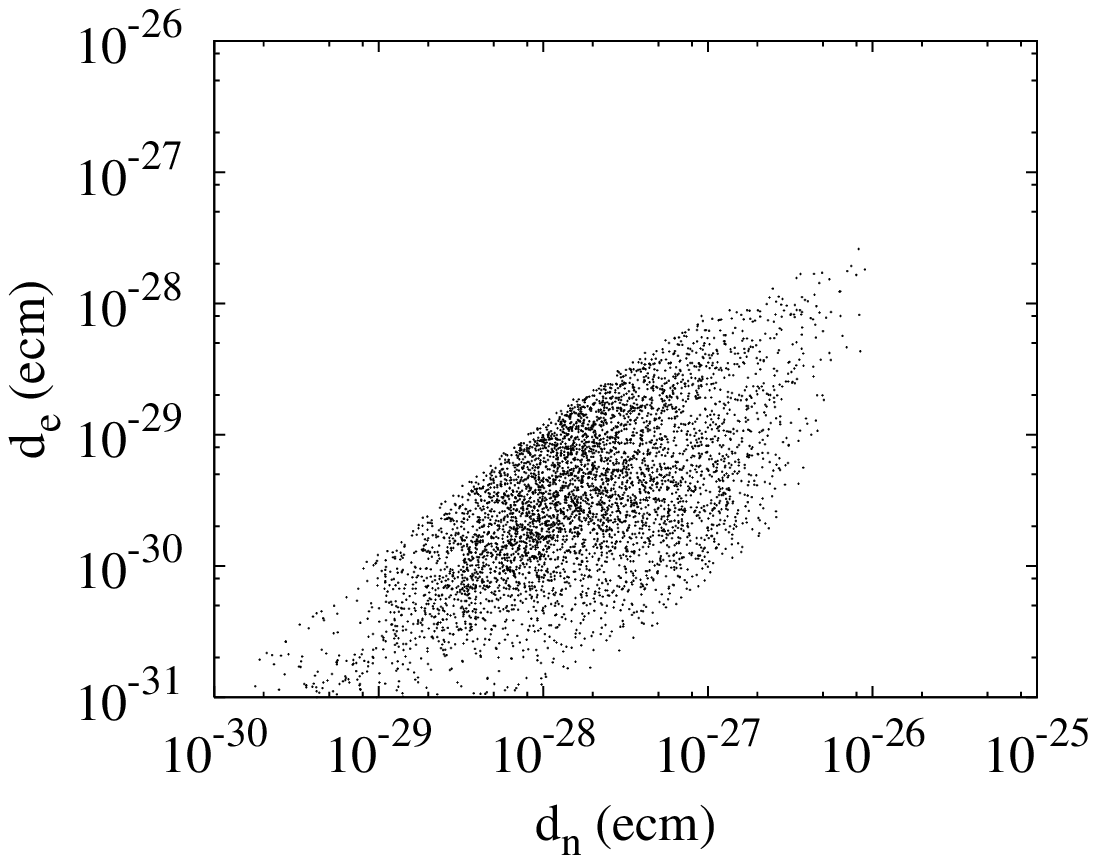} }}
\vspace*{0.3cm}
 \caption{\footnotesize 
 $d_e$ vs $d_n$ in mSUGRA with small phases, $\tan\beta=5$. Left: $\phi_\mu\not= 0$, right: $\phi_A \not=0$.}
\label{mSUGRA1} 
\end{figure}

\begin{figure}[t]
\centerline{\CenterObject{\includegraphics[width=5.5cm]{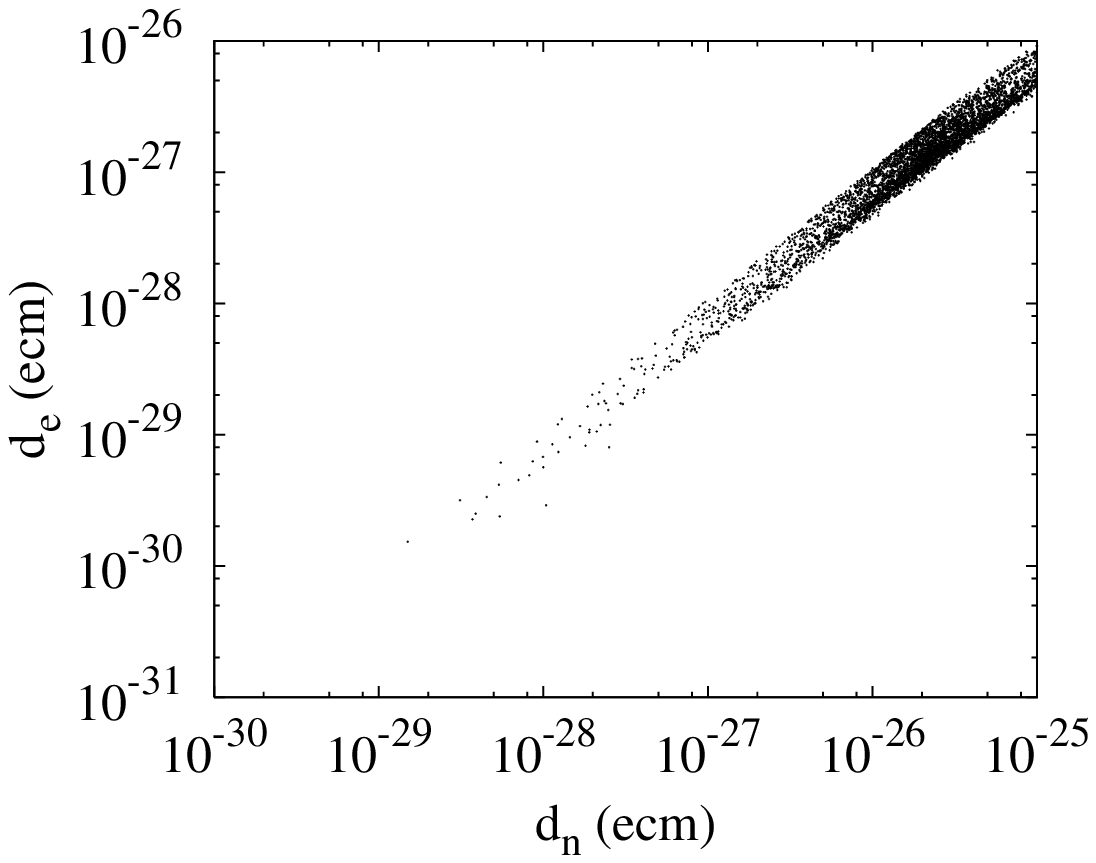} \hspace{0.5cm}
\includegraphics[width=5.5cm]{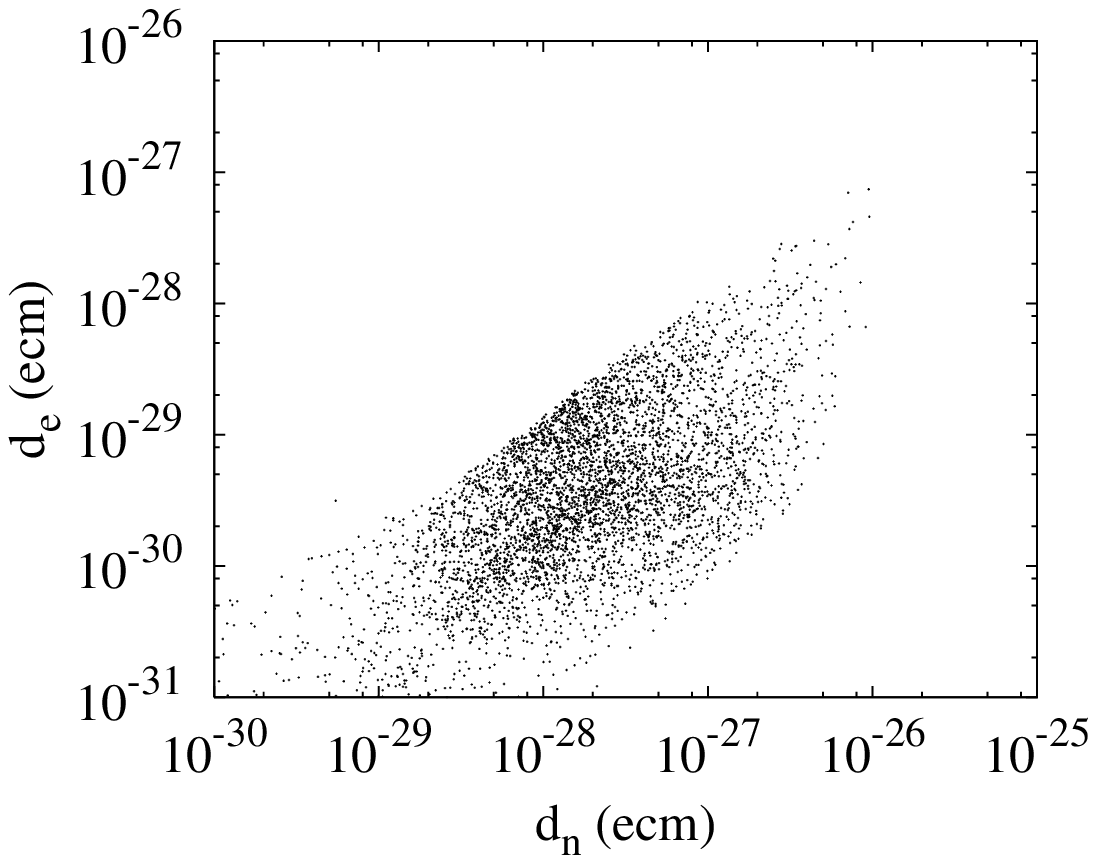} }}
\vspace*{0.3cm}
 \caption{\footnotesize As in Fig.\ref{mSUGRA1}, but for $\tan\beta=35$.
}
\label{mSUGRA2} 
\end{figure}

Let now turn to our numerical results
presented in Figs. \ref{mSUGRA1},\ref{mSUGRA2}.     
In these plots, we vary $m_0,m_{1/2},\vert A \vert $ randomly in the range
200 GeV -- 1 TeV and the phases $\phi_A$, $\phi_\mu$ in the range [$-\pi/500, \pi/500$].
In the left plot, $\phi_A$ is set to zero and, in the right plot, $\phi_\mu =0$.
For non--zero $\phi_\mu$, there is a clear linear  $d_e$--$d_n$ correlation and
$d_e $ is about an order of magnitude below $d_n$, as expected. 
$d_e$ is dominated by the chargino diagram, whereas $d_n$ receives
comparable contributions from the charginos and gluinos. 
At higher
$\tan\beta$, the EDMs increase linearly. For $\phi_A \not=0$, the correlation
is less pronounced. For a given $d_n$, the spread of $d_e$ values  is about two
orders of magnitude.
This is because $d_n$ is dominated by the gluino diagrams, whereas $d_e$ 
is dominated  by the 
neutralino contributions. The former are relatively insensitive to $m_0$ since
the squark masses are dominated by the gluino RG contributions, whereas the 
latter are sensitive to $m_0$ through both the slepton masses and the $\mu$--parameter.
Thus, fixing $d_e$ does not determine $d_n$ accurately. Note that EDMs induced by $\phi_A$ 
do not receive $\tan\beta$--enhancement.

\subsection{Heavy SUSY spectrum}

The SUSY contributions to EDMs are suppressed if the entire SUSY spectrum is in the
TeV range (Eq.(\ref{edm})). Such a possibility is motivated by the strong bound on the Higgs mass
which requires the stop mass to be of order 1 TeV. If all SUSY masses are controlled by the 
same scale, the spectrum is heavy. 
This senario can be motivated in various ways, see e.g.
 Ref.\cite{Falkowski:2005ck}.

In this class of models, the EDMs are usually dominated by the one loop diagrams of Fig.\ref{f1}
and the analysis is very similar to that presented in the previous subsection.
We note that if the $d_e$ experiments with  the $10^{-30}~ e\;{\rm cm}$ sensitivity
yield a null result, it would imply that the scale of SUSY masses is 100 TeV (Eq.(\ref{edm})).
Models with such a high SUSY breaking scale are disfavoured by the gauge coupling unification
 and naturalness considerations. Thus, again one expects a non--zero  result  in the next
round of EDM experiments.

\begin{figure}[t]
\centerline{\CenterObject{\includegraphics[width=5.5cm]{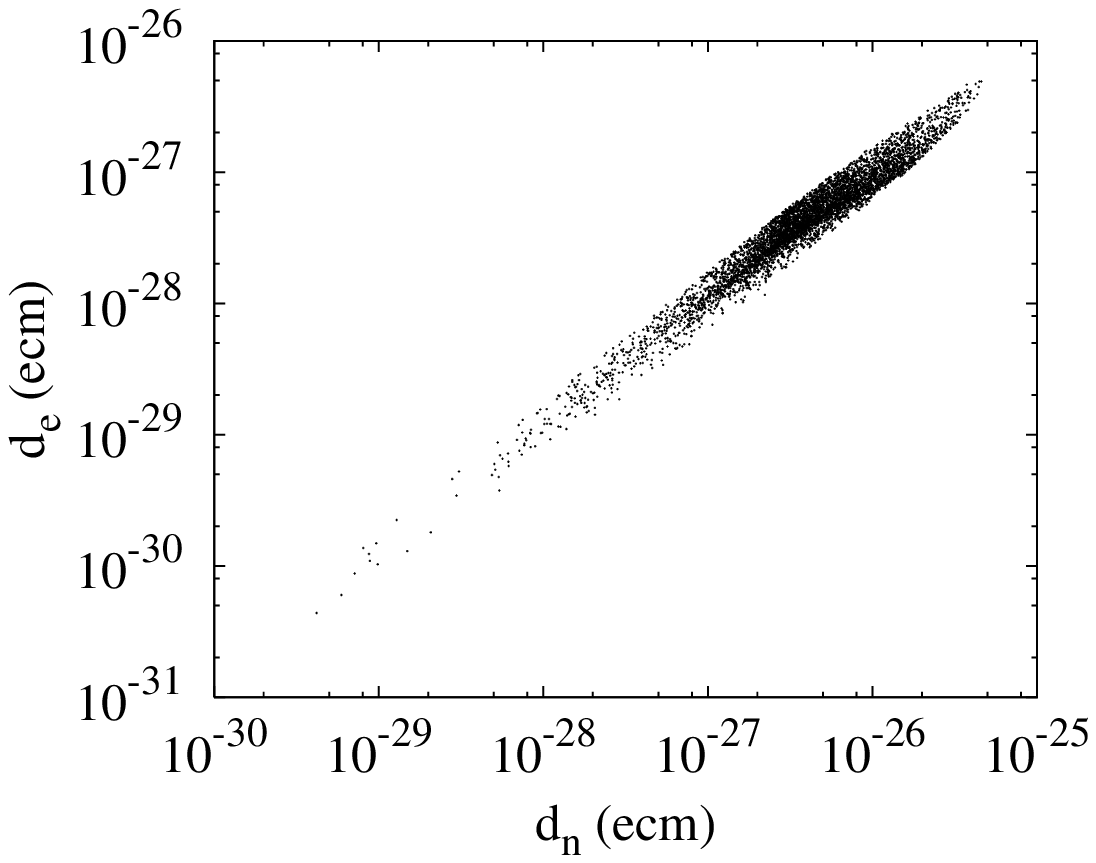} \hspace{0.5cm}
\includegraphics[width=5.5cm]{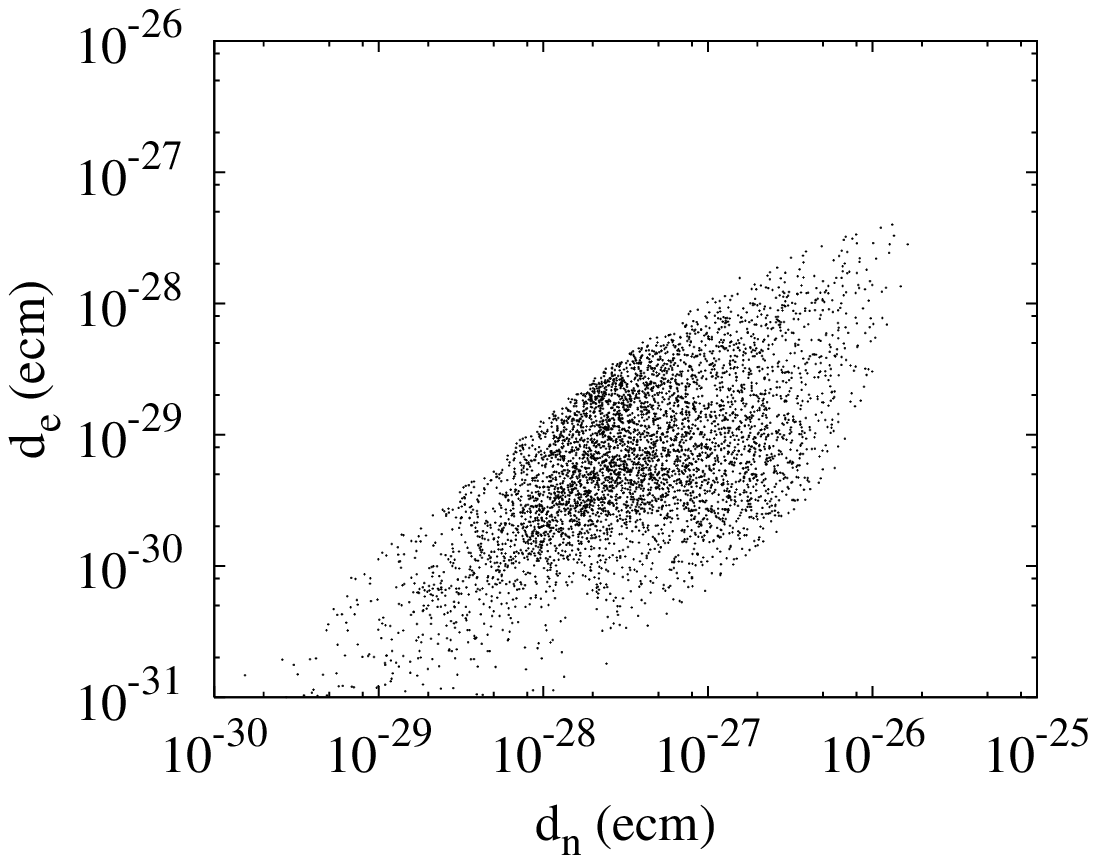} }}
\vspace*{0.3cm}
 \caption{\footnotesize 
$d_e$ vs $d_n$ in mSUGRA with a heavy spectrum, $\tan\beta=5$. Left: $\phi_\mu\not= 0$, right: $\phi_A \not=0$.
}
\label{heavy1} 
\end{figure}

\begin{figure}[t]
\centerline{\CenterObject{\includegraphics[width=5.5cm]{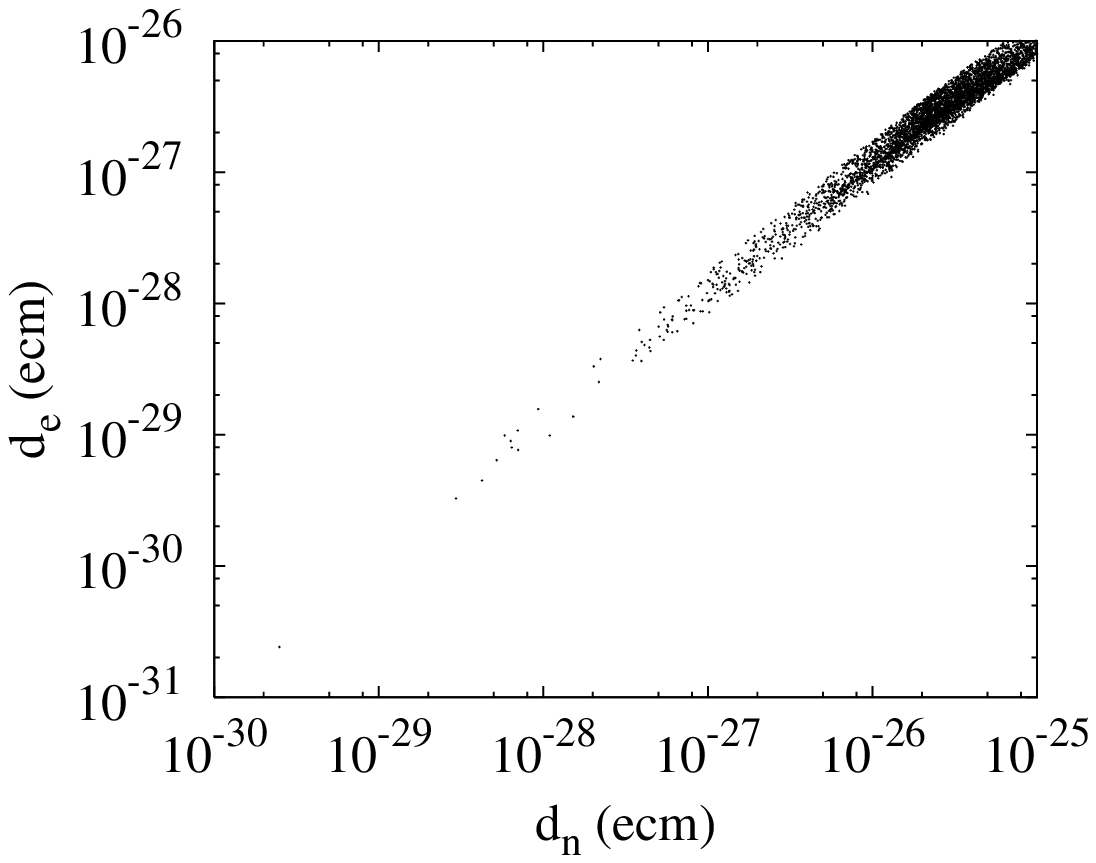} \hspace{0.5cm}
\includegraphics[width=5.5cm]{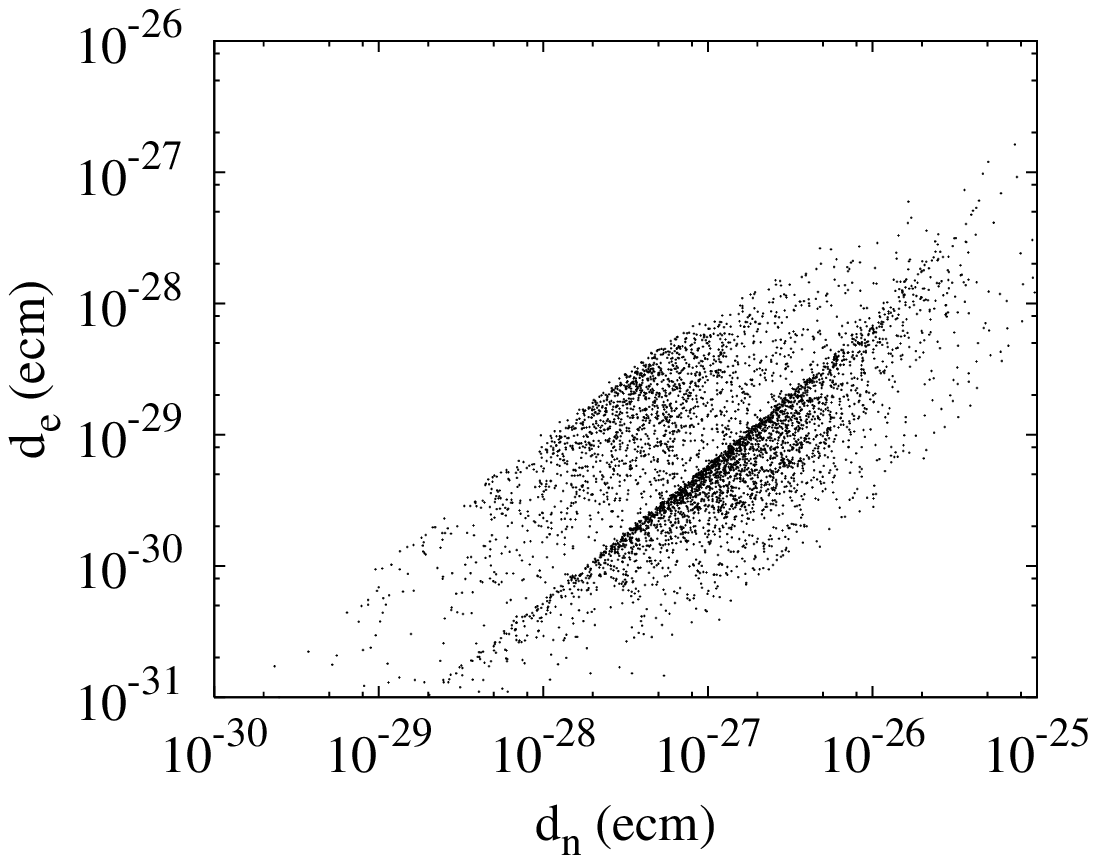} }}
\vspace*{0.3cm}
 \caption{\footnotesize 
As in Fig.\ref{heavy1}, but for $\tan\beta=35$.
}
\label{heavy22} 
\end{figure}

In our numerical analysis, we  study the mSUGRA model (Eq.(\ref{msugra})) with $m_0,m_{1/2},\vert A \vert $
in the range 2 TeV -- 10 TeV and  $\phi_A$, $\phi_\mu$  in the range [$-\pi, \pi$].
Our results are shown in Figs. \ref{heavy1},\ref{heavy22}. The  $d_n$--$d_e$   correlation
patterns  are very similar to those in the small phase scenario, namely there is a
well defined correlation in the $\phi_\mu \not=0$ case, whereas for $\phi_A \not=0$
it is far less pronounced.

\subsection{Decoupling }

The largest contributions to EDMs come from sfermions of the first two generations.
If these are very heavy, $>$ 10 TeV, the most dangerous contributions are suppressed \cite{Nath:dn}.
The third generation is required to be light by naturalness 
and contributes to EDMs at the 2 loop level.
This decoupling scenario can be  realized,
for instance, in certain types of GUT models 
where the hierarchy between the first two and third generation masses appears
due to  RG running \cite{Bagger:1999ty}.

\begin{figure}[t]
\centerline{\CenterObject{
\epsfig{figure=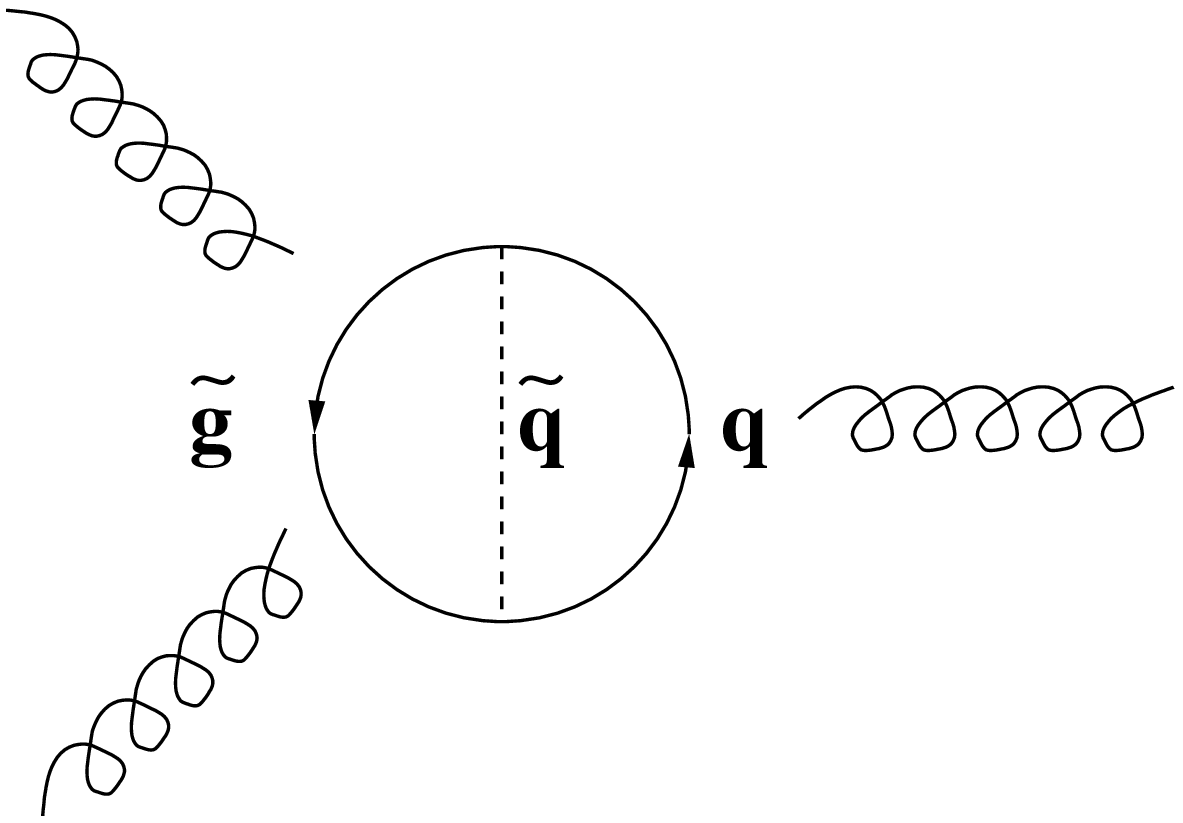,height=4cm,width=6cm,angle=0}
\hspace{1cm}
\epsfig{figure=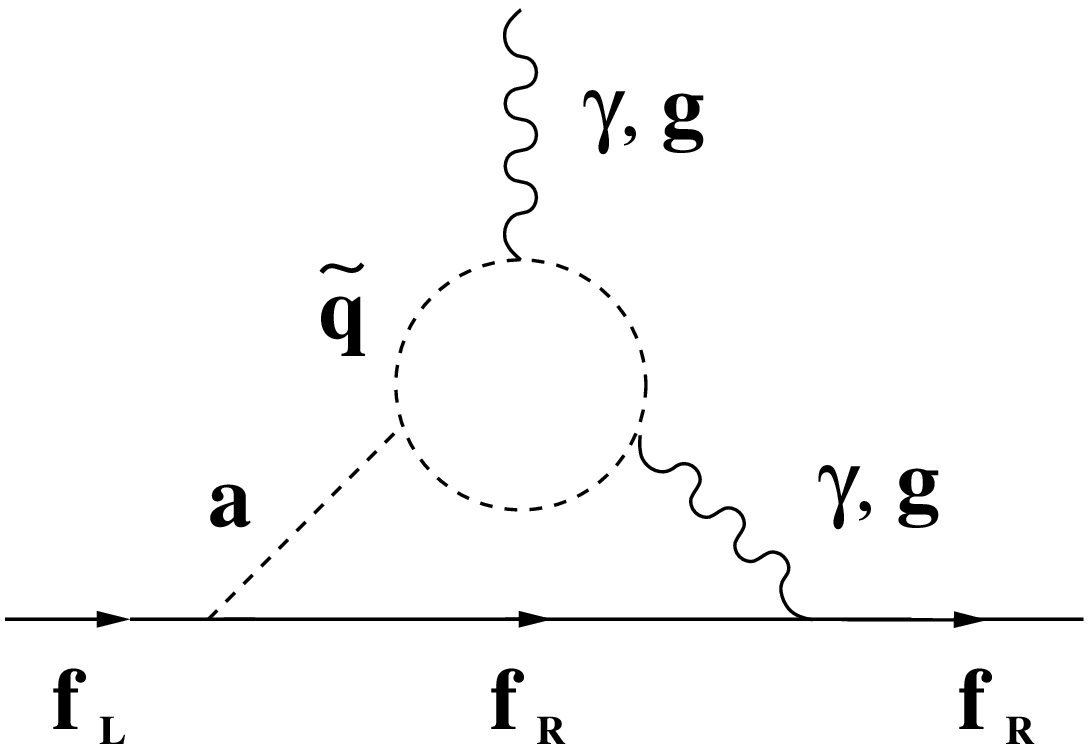,height=4cm,width=6cm,angle=0}
}}
\medskip
\caption{\footnotesize 2 loop EDM contributions.
The diagram on the left induces the Weinberg operator.
On the right, is an  example of the  Barr--Zee type diagram 
 ($a$ is a pseudoscalar Higgs).}
\label{f2}
\end{figure}

The neutron EDM is dominated  by the stop and sbottom contributions to the Weinberg
operator \cite{Dai:xh}, whereas  
the electron EDM is  due to the Barr--Zee \cite{Barr:vd}  type
2 loop diagrams \cite{Chang:1998uc}, Fig.\ref{f2}. 
This class of diagrams  also includes graphs with internal  charginos and 
$h,H,A$
Higgs bosons \cite{chang2}. Thus,
\begin{eqnarray}
&& d_n \simeq d_n (w) \;, \nonumber\\
&& d_e = d_e^{{\rm Barr-Zee} (\tilde f, \tilde \chi^+, \tilde \chi^0)} \;.
\end{eqnarray}
We note that the Barr--Zee type diagrams also contribute to the 
neutron EDM, but 
these are suppressed  compared to the 
Weinberg operator \cite{Abel:2001vy}. 
The correlation between $d_e$ and $d_n$ is rather subtle since $d_n$ depends
on the gluino and the third generation squark masses, whereas $d_e$ depends 
on the latter as well as  the chargino and Higgs masses.
An order of magnitude estimate of the resulting EDMs can be obtained by setting all SUSY masses
(apart from the first two generation sfermions) 
to be $M \gg M_Z$ and the CP phases to be given by a single quantity $\phi$. Then,
for moderate $\tan\beta$ \cite{Dai:xh} --\cite{chang2}, 
\begin{eqnarray}
&& d_n \sim \left( {300 {\rm ~GeV} \over M} \right)^2 \tan\beta ~  \sin\phi  \times 10^{-25} e\;{\rm cm}\;,\nonumber\\
&& d_e \sim \left( {300 {\rm ~GeV} \over M} \right)^2 \tan\beta~\sin\phi  \times 10^{-27} e\;{\rm cm} \;.
\label{decoup}
\end{eqnarray}
Here $d_n$ and $d_e$ differ  by almost two orders of magnitude, whereas in other scenarios
this hierarchy is usually a factor of ten.
\begin{figure}[t]
\centerline{\CenterObject{\includegraphics[width=5.5cm]{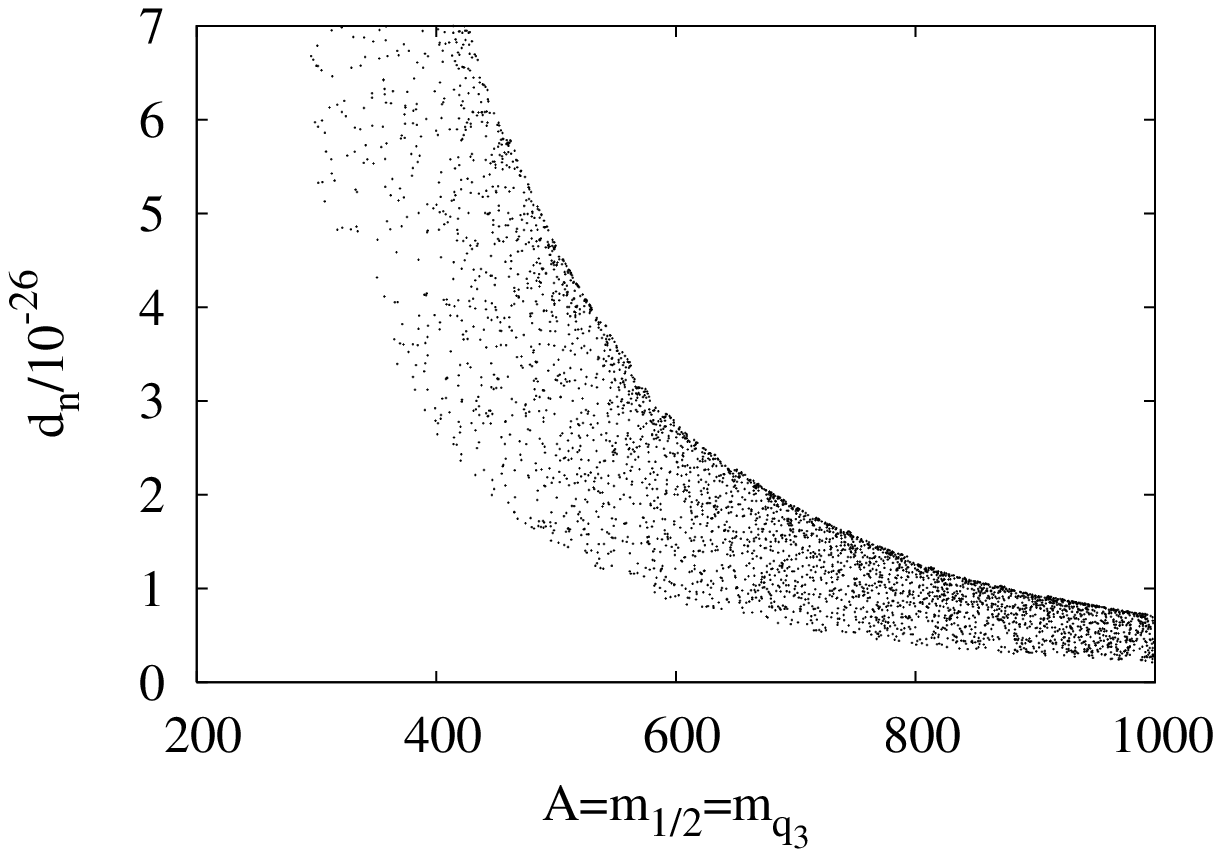} \hspace{0.5cm}
\includegraphics[width=5.5cm]{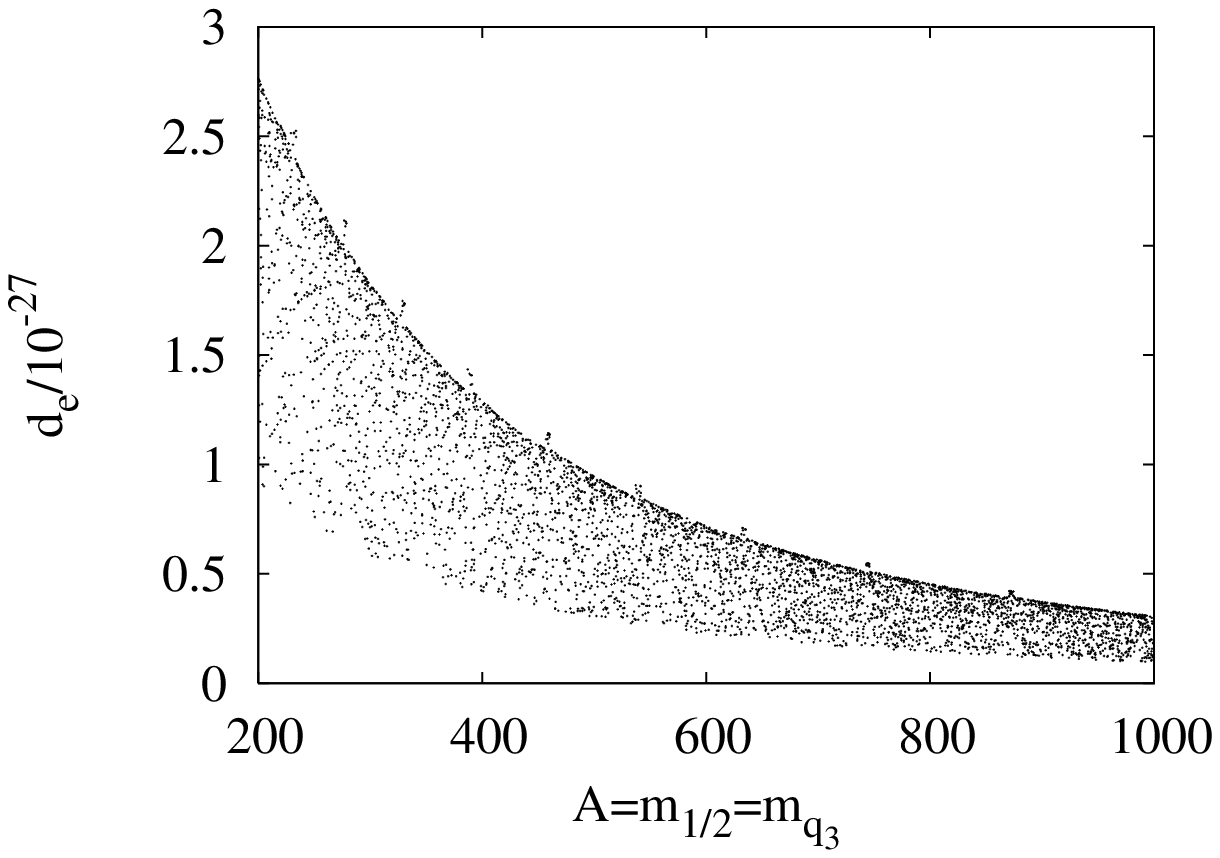} }}
\vspace*{0.3cm}
 \caption{\footnotesize 
$d_n^{\rm NDA}$ (left) and $d_e$ (right) in the decoupling scenario, 
$\phi_\mu\in [\pi/10,\pi/2]$, $\tan\beta=3$. The GUT scale parameters $A=m_{1/2}=m_{\tilde q_3}$
 are in GeV and the EDMs are  in $e$ cm.  
The  QCD sum rules model gives $d_n^{\rm SR}$  a factor of a few  smaller. }
\label{phib} 
\end{figure}

In this approximation, the mercury EDM is less sensitive to the CP phase $\phi$. This is because
the Weinberg operator contribution to $d_{\rm Hg}$ is negligible and the EDM is generated by
the subleading quark CEDMs. When $M$ or $\phi$ in Eq.(\ref{decoup}) is adjusted to satisfy
the current neutron EDM bound, the resulting $d_{\rm Hg}$ is $\leq {\cal O}(10^{-29})~e$ cm,
which is likely to be below the reach  of the mercury EDM experiments.
The deuteron EDM, however, is again  found to be similar to $d_n$.

In our numerical analysis,
we use the following GUT input parameters
\begin{eqnarray}
&~~~~~~~~~~~~~~~~~~~~m_{3}~, ~m_{1/2}~,~\vert A \vert~, ~\tan\beta~& 
\end{eqnarray}
and CP phases $\phi_A$, $\phi_\mu$. Here $m_3$ is the mass parameter for all the scalars
apart from those for the first two sfermion generations, which are assumed to be
decoupled. As in the mSUGRA case, 
$\vert \mu \vert$ and $\vert B\mu \vert$ are determined by  radiative electroweak symmetry breaking.
We vary $m_3,m_{1/2},\vert A \vert $ in the range 200 GeV -- 1 TeV and 
 $\phi_A$, $\phi_\mu$  in the range [$-\pi, \pi$].
Fig.\ref{phib} shows  typical ranges of the $d_n$ and $d_e$ values  for order one CP 
phases\footnote{ $d_n$ and $d_e$ are dominated by contributions 
sensitive to $\phi_\mu$.}.
Clearly, $d_n \sim 10^{-26}~ e\;{\rm cm} $ and $ d_e \sim 10^{-27}-10^{-28}   ~ e\;{\rm cm}$
are expected in this scenario. Smaller EDMs would imply that either the CP
phases are small or that the SUSY spectrum is heavy, including the third generation
sfermions.

\begin{figure}[t]
\centerline{\CenterObject{\includegraphics[width=5.5cm]{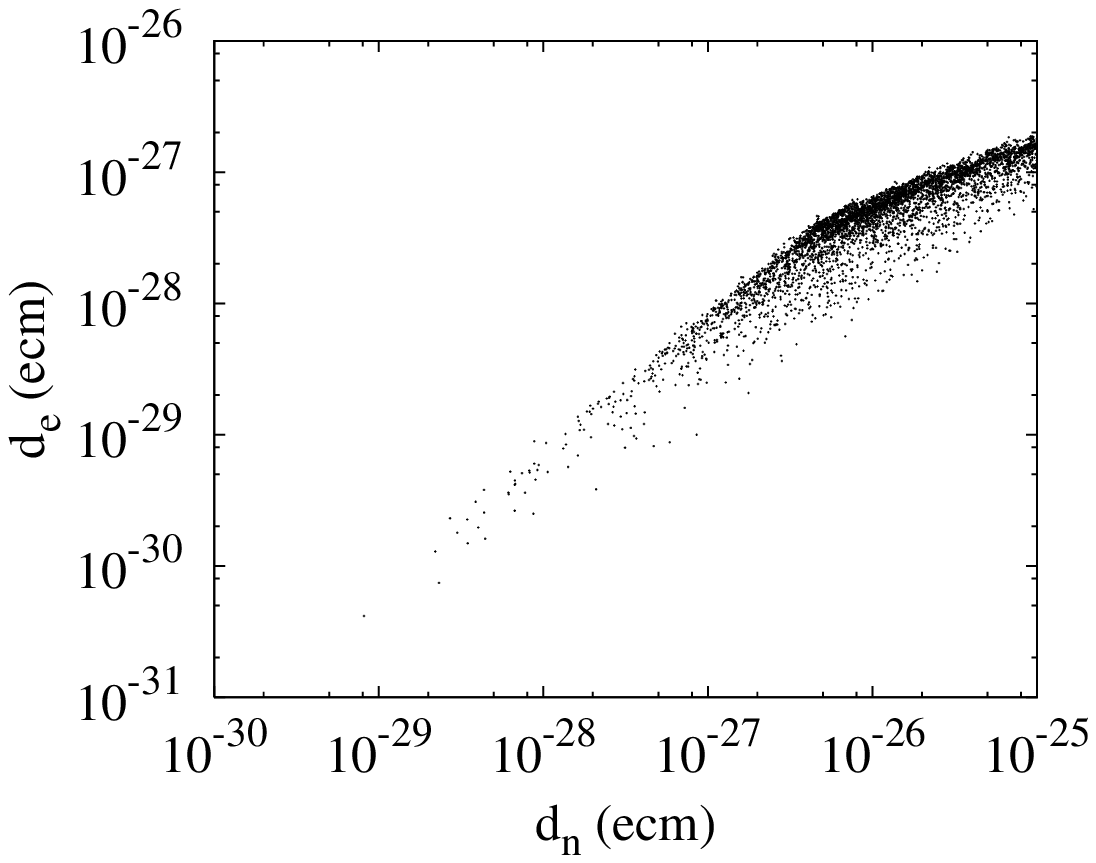} \hspace{0.5cm}
\includegraphics[width=5.5cm]{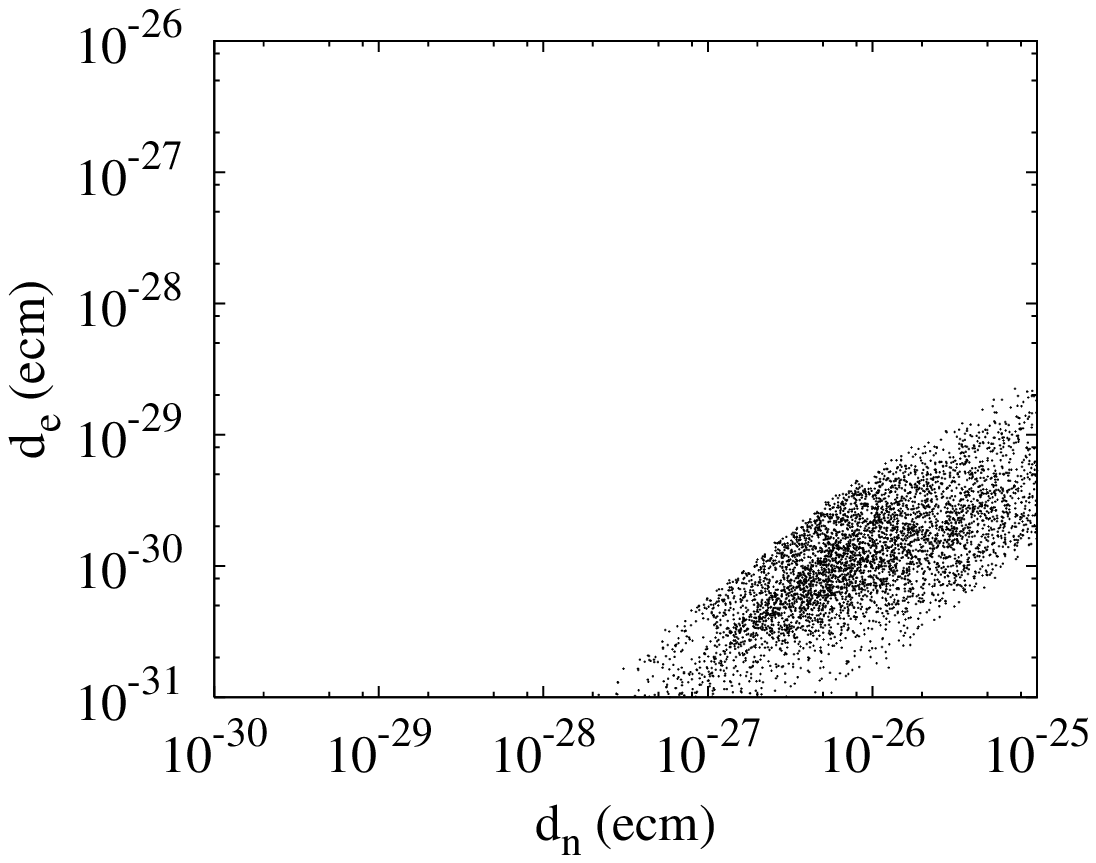} }}
\vspace*{0.3cm}
 \caption{\footnotesize 
 $d_e$ vs $d_n$ in the decoupling scenario, $\tan\beta=5$. Left: $\phi_\mu\not= 0$, right: $\phi_A \not=0$.}
\label{decoup1} 
\end{figure}

\begin{figure}[t]
\centerline{\CenterObject{\includegraphics[width=5.5cm]{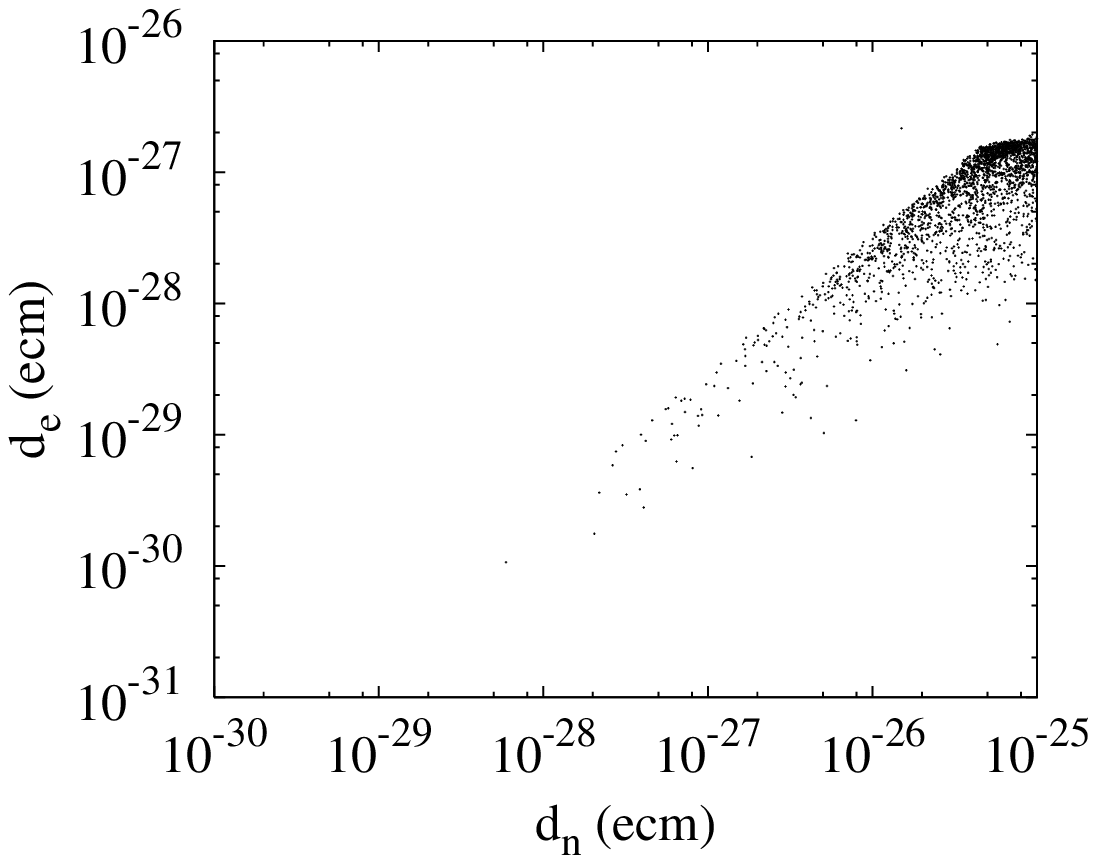} \hspace{0.5cm}
\includegraphics[width=5.5cm]{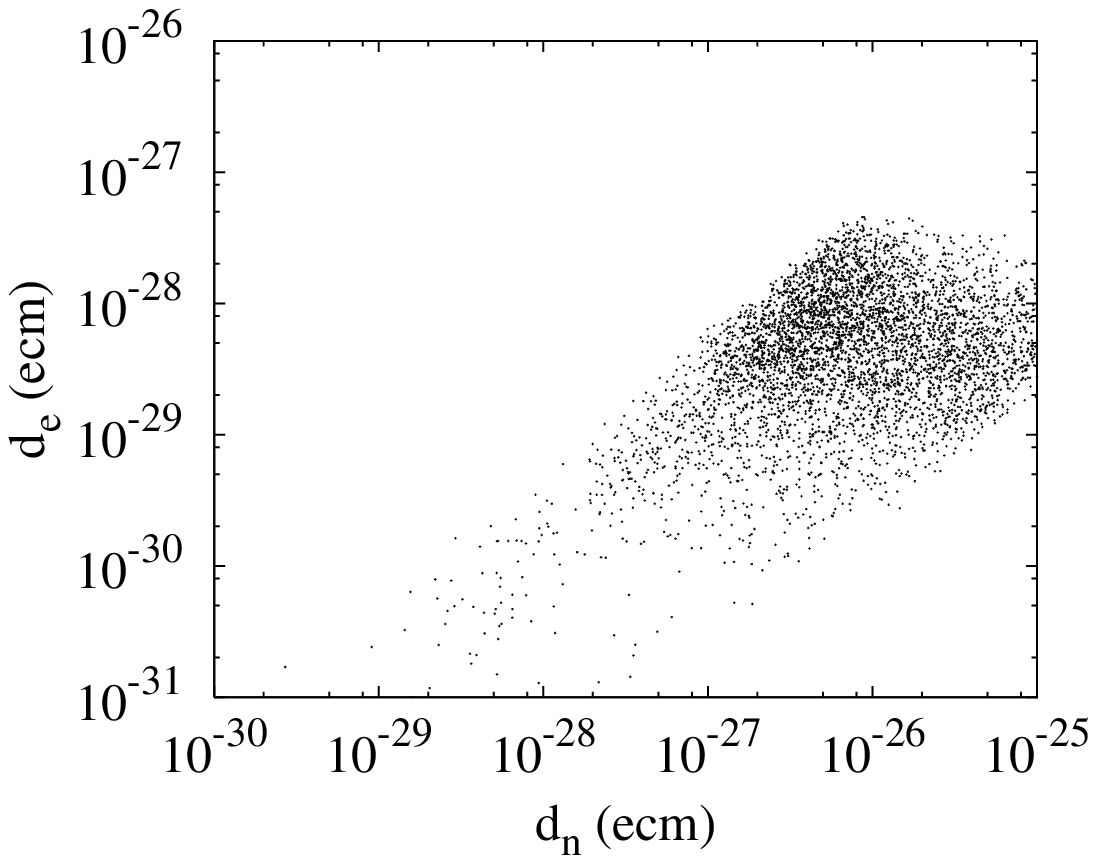} 
}}
\vspace*{0.3cm}
 \caption{\footnotesize As in Fig.\ref{decoup1}, but for $\tan\beta=35$.
}
\label{decoup2} 
\end{figure}

The corresponding correlations are presented in Figs. \ref{decoup1},\ref{decoup2}. 
For $\phi_\mu \not=0$,  $d_e$ and $d_n$ can differ
by two or one order of magnitude, depending on the  balance between the Weinberg operator
and the chargino Barr--Zee contributions. 
At larger $\tan\beta$, the situation remains the same except the EDMs increase
proportionally.
Note that since we scan over CP phases in the region [$-\pi,\pi$],
we include the possibility that the CP phases are small 
such that smaller values
of the EDMs compared to those in Fig.\ref{phib}  are allowed. 
For $\phi_A \not=0$, the hierarchy between $d_n$ and $d_e$ increases to three
or four orders of magnitude. This is because $d_n$ is dominated by the stop
contribution to the Weinberg operator, whereas the leading chargino 
Barr--Zee contribution to $d_e$ is now absent and $d_e$ is due to the 
stop Barr--Zee diagram. At larger $\tan\beta$, this hierarchy reduces since
unlike  the Weinberg operator, Barr--Zee contributions receive  $\tan\beta$--enhancement.

\subsection{Split SUSY}

This is an extreme version  of the decoupling scenario in which the
third generation sfermions are decoupled as well and naturalness is 
abandoned \cite{Arkani-Hamed:2004fb}, \cite{Arkani-Hamed:2004yi}.
From the EDM perspective, it provides an interesting framework in which the 
neutron and electron EDMs are generated by the same type of diagrams and  thus
are highly correlated.

The EDMs are induced by a version of the Barr--Zee type diagram, Fig.\ref{f2},
with the sfermion loop replaced by the chargino loop 
\cite{chang2}. Since all the Higgses
except for the SM--like $h$ are assumed to be heavy, the chargino loop 
is attached to the fermion line through $h$ and a 
photon \cite{Arkani-Hamed:2004yi}, or $W^+$ and  $W^-$ \cite{Chang:2005ac}.\footnote{
It has recently been shown \cite{Giudice:2005rz}   
that similar diagrams mediated by $h$ and $Z$
are also significant for the neutron EDM. This does not affect our
numerical estimates.}
The relevant formulae can be found in Refs.\cite{chang2},\cite{Chang:2005ac}.
Clearly, the contributing diagrams are the same for the electron and the neutron
such that  the EEDM and NEDM are  strongly correlated. 
We have therefore,
\begin{eqnarray} 
&& d_q=d_q^{{\rm Barr-Zee} (\tilde \chi^+,\tilde \chi^0 )} \;, \nonumber\\
&& d_e=d_e^{{\rm Barr-Zee} (\tilde \chi^+,\tilde \chi^0 )}\;. 
\end{eqnarray}
Simple estimates are obtained by setting the chargino mass scale $M$ to be much larger 
than the Higgs mass $\sim 100$ GeV \cite{Arkani-Hamed:2004yi},
\begin{eqnarray}
&& d_n \sim \left( {300 {\rm ~GeV} \over M} \right)^2  ~  {\sin\phi \over \tan\beta} \times 10^{-26} e\;{\rm cm}\;,\nonumber\\
&& d_e \sim \left( {300 {\rm ~GeV} \over M} \right)^2 ~{\sin\phi\over \tan\beta}  \times 10^{-27} e\;{\rm cm} \;.
\label{split}
\end{eqnarray}
Here $d_n$ is obtained  by rescaling $d_e$ with
the factor $m_q/m_e \sim 10$.
The $\tan\beta$ suppression can be traced down to the fact that the EDMs are due to the 
chargino mass matrix rephasing invariant $M_{11} M_{22} M_{12}^* M_{21}^* \propto \sin\beta\;\cos\beta$
and the SM--like Higgs couplings have no $\tan\beta$--enhancement.  

\begin{figure}[t]
\centerline{\CenterObject{\includegraphics[width=6cm]{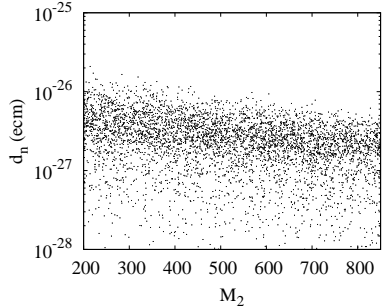}  }}
\vspace*{0.3cm}
 \caption{\footnotesize 
$d_n$ in split SUSY. Here $\phi_\mu, \phi_{M_1}$ are varied in the range [$-\pi,\pi$]; 
$M_1,  \vert \mu \vert \in $ [200 GeV, 1 TeV] and $m_h \in $ [100 GeV, 300 GeV].
}
\label{splitfigure} 
\end{figure}

We note that since no CEDMs are generated at this level, the mercury EDM is suppressed.
The deuteron EDM is, as usual, of order $d_n$.

In our numerical analysis, we vary the GUT scale gaugino masses $M_{1,2,3}$ 
in the range 200 GeV -- 1 TeV
and $\phi_\mu$ in the range [$-\pi, \pi$]. 
We set the EW scale $\mu$--parameter  by hand in the range 200 GeV -- 1 TeV and also
vary $m_h$ in the range 100 GeV -- 300 GeV. 
Typical $d_n$ values are shown in Fig.\ref{splitfigure}, whereas the corresponding
$d_e$ is found by a simple rescaling.
For order one CP phases and low $\tan\beta$, $d_n$ is between $10^{-26}$ and $10^{-27}$
$e$ cm. Smaller values are  obtained at large $\tan\beta$ or for 
small CP phases/heavy gauginos.

\begin{figure}[t]
\centerline{\CenterObject{\includegraphics[width=5.5cm]{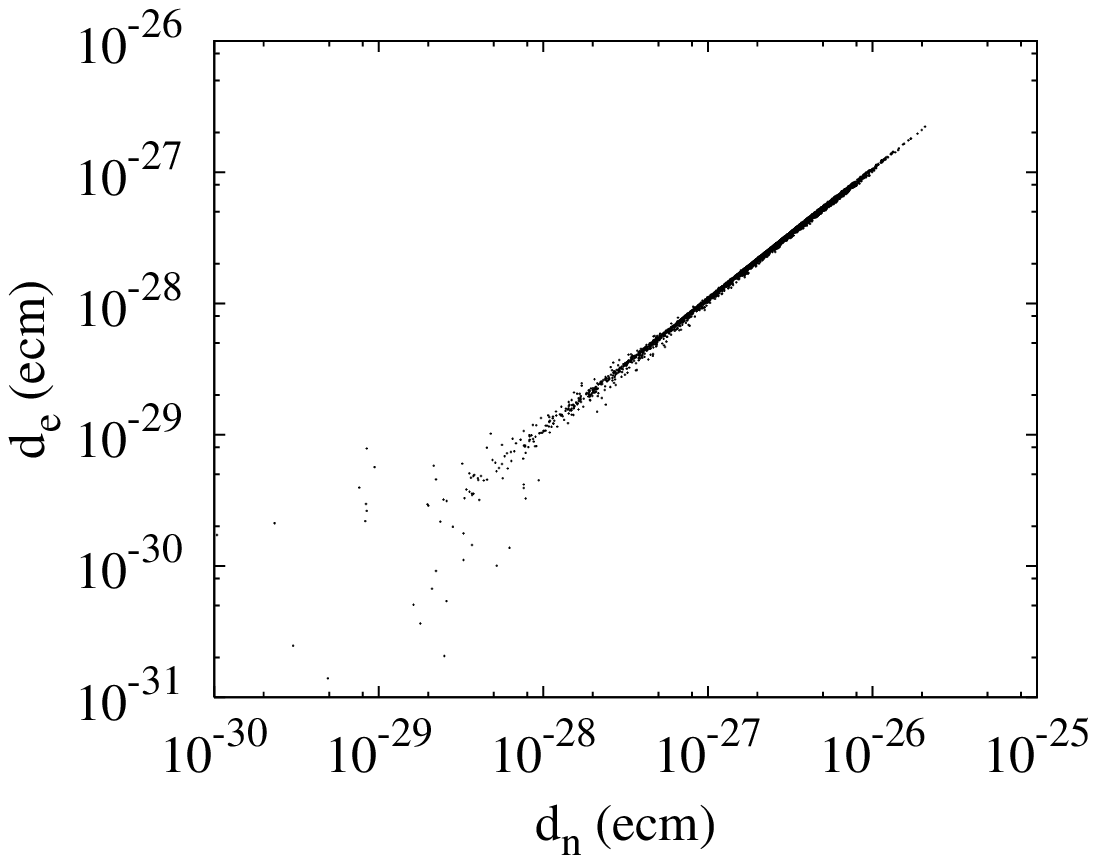} \hspace{0.5cm}
\includegraphics[width=5.5cm]{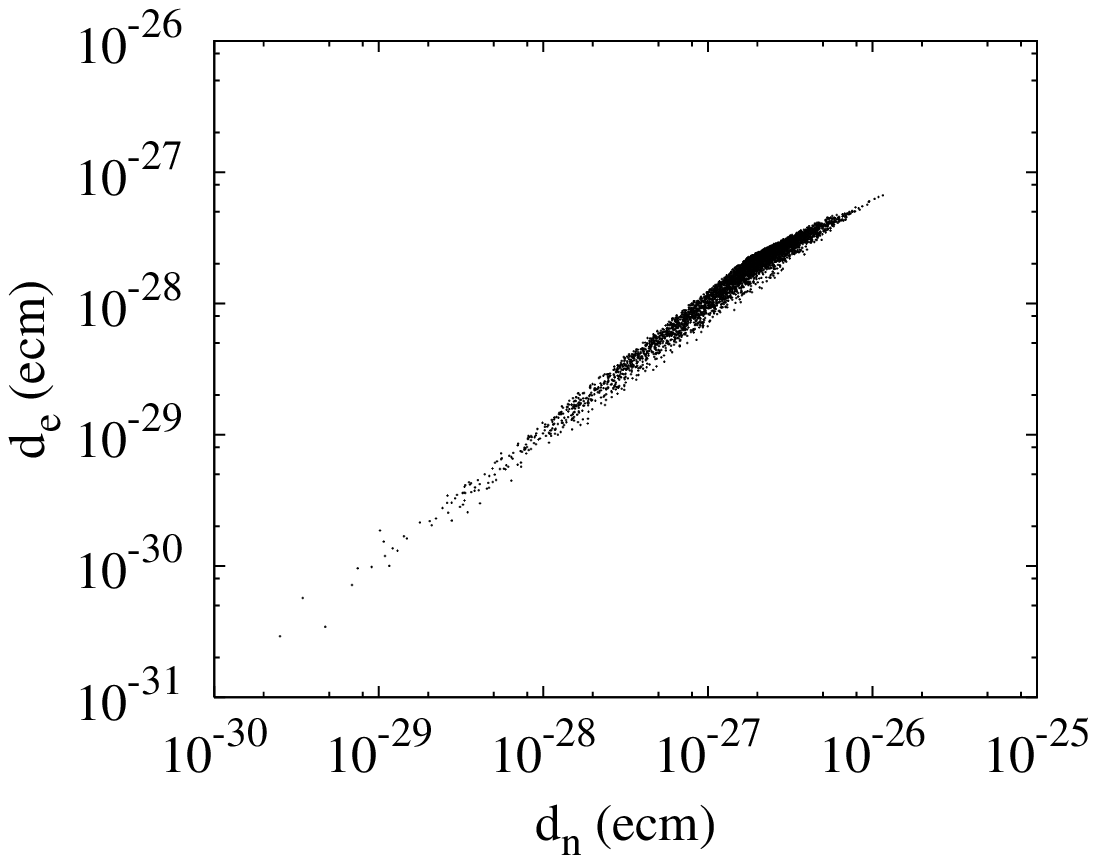} }}
\vspace*{0.3cm}
 \caption{\footnotesize 
 $d_e$ vs $d_n$ in split SUSY.  Left: $\tan\beta=5$, right: $\tan\beta=35$.}
\label{split1} 
\end{figure}

The $d_n$--$d_e$  correlations are presented in Fig.\ref{split1}. 
As expected, $d_n$ is almost  in one-to-one correspondence with $d_e$.
The small spread in the EDM values comes from an interplay between the chargino and
neutralino Barr--Zee diagrams, which contribute in different proportions
to $d_e$ and $d_n$.
In the left figure, the broad tail of $d_n$ below $10^{-28}$ $e$ cm is an
artifact of our numerical procedure 
and appears due to 1--loop contributions of heavy but not completely
decoupled sfermions (we took $m_H,m_A=10^5$ GeV and $m_{\rm sferm}=10^7$ GeV).

\section{Conclusions}

We have analyzed correlations between the electron and the neutron
EDMs in typical supersymmetric models. Unlike in the case of the $\theta$--term
induced electric dipole moments, in SUSY leptonic  EDMs are enhanced  and are likely to be observed
in the next round of experiments.

Assuming that all SUSY CP phases are order one at the GUT scale,
lower bounds on the EDMs can be obtained in the decoupling and the split SUSY
scenarios:
\begin{eqnarray}
{\rm decoupling:}~~~   &d_e& \sim  ~(10^{-1}- 10^{-2} ) ~  d_n ~ \gwig ~10^{-28}~e~ {\rm cm}~, \nonumber \\
{\rm split ~ SUSY:}~~~   &d_e& \sim ~10^{-1}~ d_n  ~\gwig~ 10^{-29}~e~ {\rm cm}~, 
\nonumber 
\end{eqnarray}
where the lower bound is saturated for a TeV range SUSY spectrum (apart from
the first two generation sfermions) in the decoupling case,
  and at large $\tan\beta$ in the case of split SUSY.  

For mSUGRA with small CP phases or a heavy spectrum, no solid lower bound can be derived.
However, non--observation of $d_e$ at the level  $10^{-30}~e~ {\rm cm}$
would imply that either the CP phases are $\lwig 10^{-5}$ or the scale 
of SUSY masses is $\gwig 100$ TeV. Both of these options make  supersymmetric models
very unappealing, so if low energy supersymmetry is indeed realized in nature,
one expects $ d_e \gwig 10^{-30}~e~ {\rm cm}$.

Concerning the $d_n$--$d_e$ correlations, the main feature of supersymmetry
is that it enhances the leptonic EDMs and we observe the following
correlations:
\begin{eqnarray}
{\rm small ~phases:} ~~~&d_e& \sim~ 10^{-1}~ d_n \;, \nonumber\\
{\rm heavy~spectrum:} ~~~&d_e& \sim ~10^{-1}~ d_n \;, \nonumber\\
{\rm decoupling:} ~~~&d_e& \sim ~(10^{-1}-10^{-2})~ d_n \;, \nonumber\\
{\rm split~ SUSY:} ~~~&d_e& \sim ~10^{-1}~ d_n \;, \nonumber
\end{eqnarray}
where we have assumed that $all$ SUSY CP phases are of the same order of magnitude.
These relations as well as correlations with $d_D$ and $d_{\rm Hg}$
 can help distinguish supersymmetry from other new physics models.

We note that these correlations are quite stable to   variations in the
GUT scale soft breaking parameters. 
As long as $d_e$ and $d_n$ are generated by similar sets of diagrams, the most
important factor for $d_e/d_n$ is $m_e/m_q$ and other effects being 
subleading.\footnote{
In particular, breaking the universality
between the GUT scale squark and slepton masses does not affect our results
significantly: the squark masses at low energies are dictated
by the gluino mass and are much less sensitive to $m_{\rm squark}^2$(GUT).}

\begin{figure}[t]
\centerline{\CenterObject{\includegraphics[width=8.0cm]{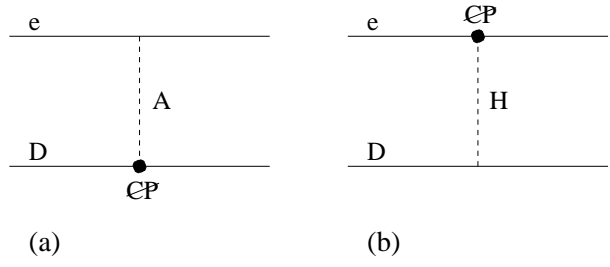}}}
\vspace*{0.3cm}
 \caption{\footnotesize 
An example of a
four--fermion operator inducing EDMs of composite objects \cite{Lebedev:2002ne}.
Here, a  pseudoscalar (a) and scalar (b)  Higgs exchange induces a CP odd  
contact interaction $\bar d d ~ \bar e i\gamma_5 e$
contributing to $d_{\rm Tl}$. 
}
\label{4f} 
\end{figure}

In this paper, we have focused on the low and moderate $\tan\beta$ regimes 
where the effects of four--fermion operators (Fig.(\ref{4f}))  are less important \cite{Barr:1991yx}
--\cite{Demir:2003js}  and have neglected certain two loop RG effects
 \cite{Olive:2005ru}. Also, we have not imposed other experimental
constraints such as the bound on the Higgs mass, etc. which are expected 
to restrict the parameter space further.
Clearly, such effects will not change the qualitative picture.

Finally, we note that in the decoupling and the small phases scenarios, at least part 
of the SUSY spectrum is light and can be observed at the LHC.
On the other hand, if the CP phases are order one,
the entire SUSY spectrum
may lie in the multi--TeV  range. Then direct discovery of the superpartners
at the LHC may prove to be quite challenging, whereas the EDMs can still probe
such a possibility.

{\bf Acknowledgements.} We are grateful to P. Harris and M. Pospelov for helpful communications.
We would also like to thank D. Doyle for collaboration at the early
stage of this project.

\newpage

\clearpage

\newpage
\clearpage

\end{document}